# Dynamics of Butane and 1-Butene in ZIF-8 Probed by Solid-State $^2$H NMR

*Alexander E. Khudozhitkov,* *,†,‡ *Sergei S. Arzumanov,*†,‡ *Daniil I. Kolokolov,*†,‡ *Alexander G. Stepanov*†,‡

† Novosibirsk State University, Pirogova Street 2, Novosibirsk 630090, Russia

‡ Boreskov Institute of Catalysis, Siberian Branch of Russian Academy of Sciences, Prospekt Akademika Lavrentieva 5, Novosibirsk 630090, Russia

## ABSTRACT

We present a detailed $^2$H NMR characterization of molecular mobility of 1-butene and n-butane propagating through the microporous ZIF-8, a zeolitic imidazolate framework renowned for its outstandingly high stability and separation selectivity of various. The experimental characterization of n-butane and 1-butene diffusivity in ZIF-8 on the molecular scale is provided for the first time. With $^2$H NMR spin relaxation analysis we have elucidated the motional mechanism for 1-butene and n-butane guests trapped within ZIF-8 framework and derived kinetic parameters for each type of motion. The characteristic times for microscopic translational diffusion and activation barriers ($E_{C4H10}$ = 34 kJ mol$^{-1}$, $E_{C4H8}$ = 32 kJ mol$^{-1}$) for n-butane and 1-butene diffusivities have been elucidated. Finally, we show that $^2$H NMR technique is capable to provide reliable information on microscopic diffusivity in the ZIF-8 MOF even for molecules with slow diffusion rates ($<10^{-14}$ m$^2$ s$^{-1}$).

1. Introduction

Production and separation of C4 hydrocarbons is a crucial task for the chemical industry. C4 olefins are highly demanded in the market. Close volatilities, size and other basic physical properties of these molecules make their separation an extremely challenging task. In fact, the current industry standard still relies on the cryogenic distillation process despite of its tremendous energy consumption.[1] Recently a major progress has been achieved in designing less energy demanding processes based on adsorption in microporous solids.[2] The adsorptive separation is a combination of equilibrium separation (difference in adsorption enthalpies) and kinetic separation (difference in diffusivity rates).[3] Regardless the way the adsorptive separation is realized it is crucial to know the diffusivity hydrocarbon mixture components within the porous material of choice.

Metal-organic frameworks (MOFs) have received significant attention in recent years due to the versatility of physicochemical properties that can be achieved in this class of material. The number of possible adsorption based application is constantly growing as new materials are being synthesized. However, among all possible applications the separation of C4 hydrocarbons over MOFs was somehow overlooked.

Chromatographic separation of C1-C5 hydrocarbons on the columns coated with MOF-5 has been reported by Munch et al.[4] It was shown that different butane isomers can be easily separated. C4 hydrocarbons are eluted in order *i*-butane/1-butene/n-butane/*trans*-2-butene/*cis*-2-

butene. The estimated diffusivities in a stationary phase at 333 K are in range $(0.75 - 2) \times 10^{-6}$ cm$^2$/s with almost identical diffusivities for n-butane and 1-butene. Gascon et al. introduced n-butane, *i*-butane, *trans*-2-butene and *cis*-2-butene in ZIF-7.[5] ZIF-7 is a representative of zeolite imidazolate frameworks constructed with benzimidazole linkers. Its windows are much smaller than windows of MOF-5 (3 Å against 12 Å), nevertheless the material is capable to adsorb the noticeable amount of molecules with size drastically exceeding the nominal aperture of the window. It is explained by the flexibility of the framework that leads to the steps and hysteresis in adsorption isotherms. The lower saturation adsorption capacity of 1-butene allows separation of this compound from the mixture. The authors show that the adsorbed amount changes with the temperature enabling the separation of the studied isomers. Single component adsorption isotherms of C4 hydrocarbons were also measured for several other MOFs.[6,7]

The tunable nature of MOFs has been demonstrated by Eum et al. Hybrid ZIF-8$_x$-90$_{100-x}$ material was shown to have a continuous change of adsorption properties with variation of composition.[8] The diffusivity of n-butane dropped by two orders of magnitude upon transition from pure ZIF-8 to pure ZIF-90 ($3 \times 10^{-12}$ to $4 \times 10^{-14}$ cm$^2$/s at 308 K). The separation selectivity of n-butane/*i*-butane mixture at the same time increased by two orders. The possibility of gradual tuning the diffusivity rate allows producing materials for membrane permeation with desired properties. However, the information regarding the diffusion of C4 hydrocarbons in MOFs is scarce. Only two reports mentioned above provide such information and none of them study the diffusion on a molecular level.

In the recent works we have reported application of an advanced $^2$H NMR relaxation analysis to uncover slow motional modes for bulky hydrocarbons inside the ZIF-8 and UiO-66 frameworks.[9,10] It was shown that this technique allows deriving diffusivities even when the diffusion is too slow for other microscopic methods such as PFG NMR and QENS. In present study we apply $^2$H NMR to clarify the mobility of n-butane and 1-butene in ZIF-8. We provide the activation barriers and characteristic times for translational diffusion.

## 2. Experimental section

**2.1. Materials.** The ZIF-8 material was synthesized similar to earlier reported procedure.[11,12] The size of obtained crystallites was around 0.2 μm with a relatively narrow size distribution. The N$_2$ adsorption measurement of the activated at 423 K material showed a BET surface area of 1350 m$^2$/g. Deuterated n-butane-$d_6$ and 1-butene-$d_8$ were used as adsorbates in this work.

**2.2. Sample Preparation.** The preparation of the sample for NMR experiments was performed in the following manner. The powder of ZIF-8 (Zn) (~0.06 g) was placed into a special glass cell of 5 mm diameter and 3 cm length. The cell was connected to the vacuum line and the material was activated at 523 K for 6 h under vacuum. After cooling the sample back to room temperature, the material was exposed to the deuterated hydrocarbon guest in the calibrated volume (58 cm$^3$) under liquid nitrogen conditions. The quantity of the adsorbed guests was regulated by the initial vapor pressure created inside the calibrated volume, to reach a loading of ~2 molecules per ZIF-8 cavity. Molecular weight of one cavity is 1365 g mol$^{-1}$, therefore we had to create a pressure of 37 mbar of the target guest hydrocarbon inside the calibrate volume. After adsorption, the neck of the tube was sealed off, while the material sample was maintained in liquid nitrogen to prevent its heating by the flame. Prior to NMR investigations all sealed samples were kept at 373 K for 72 h to allow even redistribution of the guest molecules inside the porous material.

**2.3. NMR Measurements.** $^2$H NMR experiments were performed at the Larmor frequency $\omega_z/2\pi$ = 61.42 MHz on a Bruker Avance-400 spectrometer, using a high probe with 5-mm horizontal solenoid coil. All $^2$H NMR spectra were obtained by Fourier transformation of quadrature-detected phase-cycled quadrupole echoes.[13] Inversion-recovery experiments for the determination of the spin-lattice relaxation time ($T_1$) were performed using the pulse sequence $(180°_x – \tau_v – 90°_{\pm x} –$ acquisition $– t)$, where $\tau_v$ was a variable delay between 180° and 90° pulses, $t$ is a repetition delay of the sequence during the accumulation of the NMR signal. The duration of 90° pulse was 1.65 μs. Spin-spin relaxation time ($T_2$) was obtained by a Carr-Purcell-Meiboom-Gill (CPMG)[14] pulse sequence.

The temperature of the samples was controlled with a nitrogen gas flow at low temperatures and air flow at elevated temperatures, stabilized with a variable-temperature unit BVT-3000 with the precision of about 1 K.

**2.4. Modeling.** Modeling $^2$H NMR spectra line shape and spin relaxation rates has been performed with a homemade Fortran program based on the standard formalism applied for description of molecular motions.[15,16]

**3. Results**

Following our previous work on mobility of aromatic hydrocarbons in ZIF-8, we start first with the line shape analysis.[17] $^2$H NMR line shapes for both n-butane-$d_6$ and 1-butene-$d_8$ adsorbed on ZIF-8 represent a liquid-like narrow signal (Figure 1A,B) within the whole experimental temperature range studied (113 K – 527 K). The isotropic patterns imply that both n-butane and 1-butene guests exhibit fast isotropic rotation when confined inside the ZIF-8 cavities.

In case of selectively deuterated n-butane the spectrum consists of a single component corresponding to methyl groups (Figure 1A), which line width monotonously decreases with temperature increase as the isotropic tumbling gets gradually faster. The similar behavior shows the spectrum of 1-butene (Figure 1B), except the molecule was not selectively deuterated. Therefore the spectrum is comprised of three signals related to each functional group. However, up to 393 K spectrum remains unresolved and only above this temperature the signal of methyl group can be distinguished. Note that both $T_1$ and $T_2$ relaxations do not resolve individual signals, hence the relaxation times for all groups can be considered to be similar. For these reasons, during simulations of $T_1$, $T_2$ relaxation we regard only the methyl group dynamics.

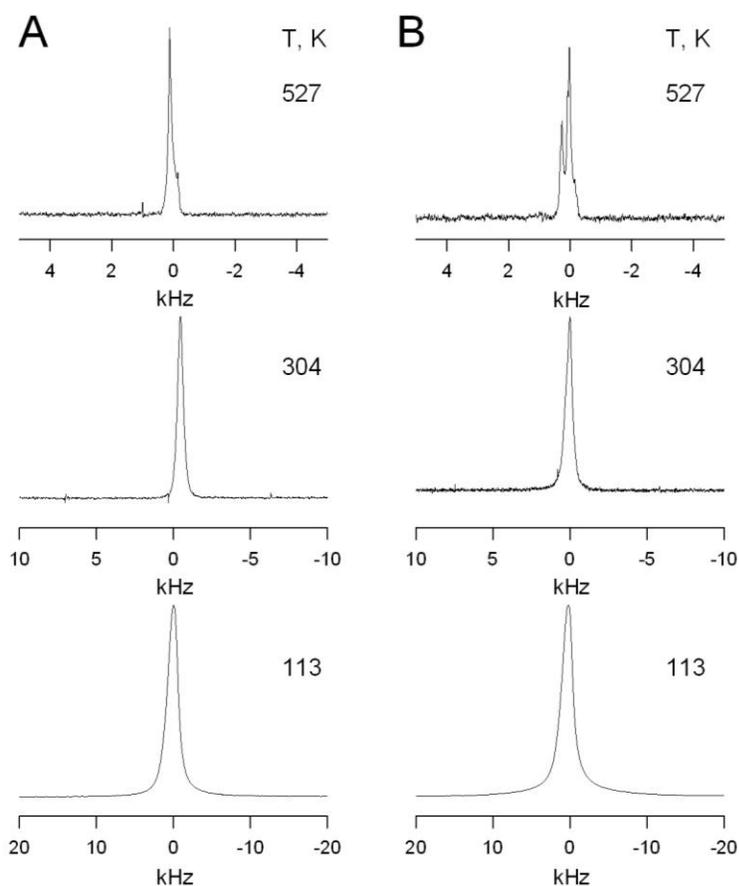

**Figure 1**. Experimental $^2$H NMR line shapes for n-butane-$d_6$ (A) and 1-butene-$d_8$ (B).

**²H NMR spin relaxation analysis.** Experimental results of the $T_1$, $T_2$ temperature dependences for n-butane and 1-butene are shown in Figures 2A and 2B. Both species exhibit qualitatively similar behavior. The $T_1$ relaxation curve grows steadily with a single slope, governed thus by a single fast motion within the temperature range studied. $T_2$ relaxation shows a remarkably unusual pattern with two ranges of temperature change. At the lowest temperature range $T_2$ remains almost unchanged. At higher temperature this pattern changes to a well-type behavior: a pronounced decrease followed by a steep growth (see, temperature areas indicated as **a** and **a'** in Figure 2). This behavior is characterized by two main points: (1) despite the line shape corresponds to isotropic rotation and the $T_1$ relaxation should be governed by a single rotation mode in a fast regime ($\tau_C \ll \omega_z^{-1}$), the $T_2$ relaxation does not coincide with $T_1$. (2) Instead of a monotonous growth, expected for the case of isotropic rotation, the $T_2$ curve shows an area of a decrease with a pronounced minimum.

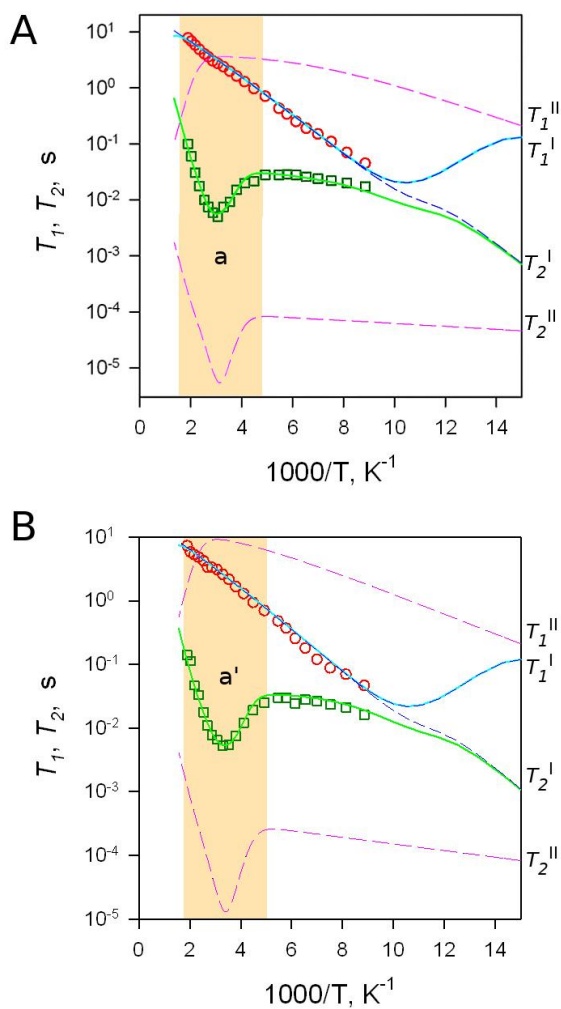

**Figure 2.** Experimental $T_1$ (○), $T_2$ (□) relaxation curves with numerical simulation results for n-butane-$d_6$ (A) and 1-butene-$d_8$ (B). Numerical simulations with elaborated model of the motion are presented in lines: individual $T_1^I$, $T_2^I$ for the state I (blue dashed); individual $T_1^{II}$, $T_2^{II}$ for the state II (pink dashed); effective $T_1$, $T_2$ after exchange (solid lines).

We have demonstrated earlier in the case of aromatic guests and propane/propene in ZIF-8[17] that such behavior can be rationalized by assuming a slow chemical exchange of guesting species between two states **I** and **II** of the guest with different localization within the framework of ZIF-8 cage and exhibiting different local mobility at each of the states.

**Model of n-butane and 1-butene motion in ZIF-8.** To rationalize the evolution of $T_1$ and $T_2$ with temperature we assume that three main motions for adsorbed hydrocarbon exist in ZIF-8 framework (Figure 4):[17] (i) free reorientation of a guest molecule at the central part of the cage (dynamic state **I**); (ii) restricted dynamics of the molecule bound to the cage wall (dynamic state **II**); (iii) the exchange between the states **I** and **II**. The exchange process represents either the exchange between the bound state **II** and the molecule in the center of the cage in the state **I** or the exchange between the state **II** and the state **I** of the molecule in the neighbor cage. The latter process represents the elementary step of the translational diffusion from one cage to another cage by squeezing through the window with a subsequent release to the neighbor cage. The exchange step of the molecule to the neighbor cage may occur through any of six windows. Consequently, when the exchange gets fast enough this motion is regarded by the deuteron spins as effectively isotropic, i.e. the isotropic self-diffusion within the framework.

In order to apply such a scheme to fit the experimental relaxation data for propane and propene we need to define the motional modes for each of the states. For the state **I** the primary motion is a fast isotropic reorientation of the molecule within the cage with characteristic time $\tau_{iso}$. In addition the uniaxial rotation of the methyl group with the time $\tau_{C3}$ was taken into account. These motional modes describe the experimental $T_1$ relaxation curves.

In the state **II**, the molecules are assumed to be relatively tightly bound to the framework walls. Only the restricted librations or local uniaxial rotations are geometrically possible.[17-19] In order to adequately describe the experimental $T_2$ relaxation curves, we have to assume the presence of some anisotropic motions with characteristic times of $\sim 10^{-7}$ s. Yet, since the exchange process affects both $T_1$ and $T_2$ curves simultaneously; one slow anisotropic motion in the state **II** is not enough to describe the $T_1$ relaxation (Figure 3B). Hence, we assume that along the slow librations the molecules in the bound state **II** exhibit also some fast anisotropic rotations. The presence of two distinct types of anisotropic motions for molecules bound to the frameworks walls is actually not surprising for ZIF-8. Recent studies have demonstrated that ZIF-8 linkers show the presence of both slow and fast "breathing" modes in the cage.[6] As we have shown earlier,[17] the guests in the bound state might follow these framework librations. The exact libration modes and their rates can be dependent on the guest type. On the other hand, for both propane and propene in the bound state we expect some local anisotropic motions. These motions can be uniaxial rotation of the methyl group and rotation of the molecule as whole around their hydrocarbon chain axis. Within our model we take these motions into account by applying the model of two independent in-cone librational motions with the times $\tau_{lib1}$ and $\tau_{lib2}$. Based on this simple scheme the individual relaxation times ($T_1^I$ and $T_1^{II}$) and ($T_2^I$ and $T_2^{II}$) in the states **I** and **II** for propane and propene are computed.

The exchange between the states **I** and **II** is described by the exchange rates $\tau_{12}^{-1}$ and $\tau_{21}^{-1}$, the corresponding equilibrium constant $K_{eq}$ and the relative populations $p_{II}$ and $p_I$. Here $\tau_{12}^{-1} = \tau_{21}^{-1} \times K_{eq}$, and $K_{eq} = p_{II}/p_I$. If we know the exchange rates, the effective relaxation times $T_1$ and $T_2$ can be computed directly from the Bloch equations as the corresponding Eigen values of the respective exchange matrices.[17,20]

The exchange rate $\tau_{12}^{-1}$ is essentially the life time of the bound state **II.** The pronounced well-type behavior of the $T_2$ curve in the high temperature range (temperature areas marked as **a** and **a'** in Figure 2) is the direct manifestation of this exchange. The exchange can be related with the jump diffusion process. Consequently, this peculiar pattern of the $T_2$ curve provides substantial information needed to extract the characteristic times and the activation barrier of this exchange motion. In fact, the exchange process appears to be more complex. Our attempt to fit the experimental $T_2$ curve with only one activation-type diffusional motion with characteristic time $\tau_D$ fails to reproduce the lower temperature range of relaxation behavior. The diffusional process occurs simply too slow to affect the $T_2$ relaxation at low temperature. To reach an adequate agreement with the experimental observations of $T_2$ evolution our model requires introduction of one additional parallel exchange process, characterized by the time $\tau_{ex}$, that would be present even at lowest temperatures, i.e. occurring with $\tau_{ex}^{-1} > 10^2$ Hz at T ~ 100 K. This second exchange process must be relatively slow ($\tau_{ex}^{-1} < 10^6$ Hz) and almost barrierless. Physical nature of this exchange process is discussed in the Discussion section.

Despite such complications, the presented above exchange scheme is a typical way to introduce the translational jump diffusion in the ordered microporous solids.[21-24] In our model, we thus introduce the exchange between the states **I** and **II** by an effective exchange rate composed of two parallel processes with $\tau_{21}^{-1} = \tau_D^{-1} + \tau_{ex}^{-1}$, where $\tau_D$ characterizes the usual activation-type jump process and $\tau_{ex}$ is in charge for second process with low activation barrier. We always assume that the correlation times $\tau$ for each individual motional process follow the Arrhenius law, i.e. $\tau = \tau_0 \exp(E/RT)$. Details on spin relaxation times computation for different cases including the presence of slow anisotropic motions are given in our previous works.[10,20]

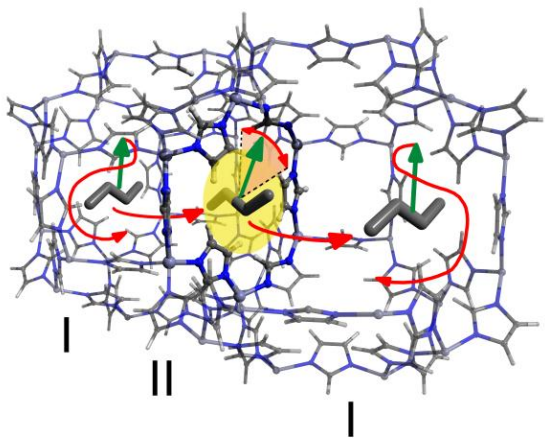

**Figure 4.** The butane (butene) guests can migrate between two dynamically different states **I** and **II** within ZIF-8 framework. The state **I** corresponds to relatively free motion at the central part of the cage. The state **II** corresponds to the motion of the guest confined near the cage wall.

Simulation results shown in Figure 2 evidence that the model discussed above gives a very good description of the experimental data for relaxation times evolution with temperature for both n-butane and 1-butene. Remarkably, the unusual behavior of the spin-spin ($T_2$) relaxation time curve is perfectly reproduced in both cases, yielding thus information on translational diffusion characteristic time $\tau_D$, including activation energies $E_D$ for diffusion. The whole set of dynamical parameters used for simulation of experimental relaxation curves is given in Table 1.

**Table 1.** Fitting parameters for spin relaxation times of n-butane and 1-butene derived from the relaxation analysis.

|  | Butane | 1-butene |
|---|---|---|
| $E_{iso}$, kJ mol$^{-1}$ | 6.8 | 6.5 |
| $\tau_{iso0}$, s | 3×10$^{-13}$ | 4×10$^{-13}$ |
| $E_D$, kJ mol$^{-1}$ | 34 | 32 |
| $\tau_{D0}$, s | 8×10$^{-11}$ | 6×10$^{-11}$ |
| $E_{C3}$, kJ mol$^{-1}$ | 3 | 3 |
| $\tau_{C30}$, s | 3×10$^{-14}$ | 3×10$^{-14}$ |
| $E_{lib1}$, kJ mol$^{-1}$ | 0.5 | 1 |
| $\tau_{lib10}$, s | 4×10$^{-7}$ | 4×10$^{-7}$ |
| $E_{ex}$, kJ mol$^{-1}$ | 1.5 | 1 |
| $\tau_{ex0}$, s | 8×10$^{-6}$ | 3x10$^{-5}$ |

The estimated accuracy is 10% for all activation barriers and 20% for all pre-exponential factors.

## 4. Discussion

All motional parameters for n-butane and 1-butene are almost identical. If close values of activation barriers for isotropic rotation inside the cavity are understandable, close values of diffusion activation barriers are somewhat surprising. In case of propane and propene in ZIF-8 the difference in diffusion activation barriers was significant (38 kJ mol$^{-1}$ vs. 13 kJ mol$^{-1}$). Apparently even one additional link in hydrocarbon chain is enough to deprive olefin of advantage in window passing process compared to paraffin. We assume that due to the double bond propylene is more rigid molecule and therefore has fewer conformations in the vicinity of the window. In case of longer molecules such as n-butane and 1-butene the conformational freedom achieved by longer hydrocarbon chain is already sufficient to impede the window passing. Interestingly, the activation barrier of n-butane diffusion is slightly smaller than the barrier of propane (34 kJ mol$^{-1}$ vs. 38 kJ mol$^{-1}$). Moreover, if the chain link is attached to alpha carbon of propane (*i*-butane) instead of beta carbon (n-butane) then the decrease of diffusion activation barrier is more pronounced (17 kJ mol$^{-1}$ for *i*-butane vs. 34 kJ mol$^{-1}$ for n-butane).[ref] So the diffusion activation barrier has the opposite size dependence in row propane, n-butane, *i*-butane.[25]

It is of interest to compare the kinetic parameters estimated for diffusivity of n-butane and *i*-butane with values previously reported by Eum et al. The $^2$H NMR provides directly only the correlation time of the motion and its activation barrier. Hence, to make the comparison with the existing data, we compute the diffusion rates based on the correlation time $\tau_D$ using the basic equation for the isotropic diffusion model (Einstein equation): $D = <l^2>/6\tau_D$, where $l$ is the average jump length corresponding to the distance between the ZIF-8 cage centers, ~ 1 nm. The diffusivity at 308 K derived from the gravimetric uptake curves by Eum et al. is about $2\times10^{-9}$ cm$^2$/s for *i*-butane and $2\times10^{-12}$ cm$^2$/s for n-butane. These values are a little bit smaller to diffusivities assessed from our data: $7\times10^{-9}$ cm$^2$/s for *i*-butane and $3\times10^{-11}$ cm$^2$/s for n-butane. However, relatively good agreement between these methods indicates that the diffusivity operates on nanometer, micrometer and macroscopic length scales following the similar dynamical mechanisms. The diffusion of 1-butene in ZIF-8 has not been studied before, so there is no data to compare with.

**Conclusion**

The mobility of 1-butene and n-butane confined in ZIF-8 MOF was characterized by $^2$H solid state NMR spectroscopy. Detailed analysis of the $^2$H NMR spin relaxation times allows building up the dynamical model of butane and 1-butene propagation through the ZIF-8 framework. In particular, it allows giving the estimations of the rates and activation barriers for the translational diffusion on a molecular level. This is the first microscopic experimental characterization of 1-butene and n-butane diffusion in ZIF-8. Activation barriers for the intercage diffusion ($E_{C4H10}$ = 34 kJ mol$^{-1}$ for n-butane and $E_{C4H8}$ = 32 kJ mol$^{-1}$ for 1-butene) were derived from relaxation analysis. The rate of diffusion assessed from our data is a good agreement with values obtained from gravimetric uptake.[8] This implies that the $^2$H NMR technique is capable to provide an information on microscopic diffusivity in the ZIF-8 MOF for the molecules with slow diffusivity (<10$^{-14}$ m$^2$/s).

**ASSOCIATED CONTENT**

**AUTHOR INFORMATION**

**Corresponding Authors**

*E-mail: a.khudozhitkov@g.nsu.ru (A.E.K.).

**Author Contributions**

The manuscript was written through contributions of all authors.

**Notes**

The authors declare no competing financial interests.

## ACKNOWLEDGMENT

This work was supported by Russian Foundation for Basic Research (18-33-00048).